# Limits of feedback control in bacterial chemotaxis

Yann S. Dufour [*,a], Xiongfei Fu [*,a], Luis Hernandez-Nunez [a,c], Thierry Emonet [a,b]

Accepted for publication at PLoS Computational Biology

[a] Department of Molecular, Cellular, and Developmental Biology, Yale University, New Haven, CT 06520
[b] Department of Physics, Yale University, New Haven, CT 06520
[c] current address: Systems Biology, Havard University, Cambridge, MA 02138
[*] these authors contributed equally. Corresponding author: thierry.emonet@yale.edu

**Abstract.** Inputs to signaling pathways can have complex statistics that depend on the environment and on the behavioral response to previous stimuli. Such behavioral feedback is particularly important in navigation. Successful navigation relies on proper coupling between sensors, which gather information during motion, and actuators, which control behavior. Because reorientation conditions future inputs, behavioral feedback can place sensors and actuators in an operational regime different from the resting state. How then can organisms maintain proper information transfer through the pathway while navigating diverse environments? In bacterial chemotaxis, robust performance is often attributed to the zero integral feedback control of the sensor, which guarantees that activity returns to resting state when the input remains constant. While this property provides sensitivity over a wide range of signal intensities, it remains unclear how other parameters such as adaptation rate and adapted activity affect chemotactic performance, especially when considering that the swimming behavior of the cell determines the input signal. We examine this issue using analytical models and simulations that incorporate recent experimental evidences about behavioral feedback and flagellar motor adaptation. By focusing on how sensory information carried by the response regulator is best utilized by the motor, we identify an operational regime that maximizes drift velocity along chemical concentration gradients for a wide range of environments and sensor adaptation rates. This optimal regime is outside the dynamic range of the motor response, but maximizes the contrast between run duration up and down gradients. In steep gradients, the feedback from chemotactic drift can push the system through a bifurcation. This creates a non-chemotactic state that traps cells unless the motor is allowed to adapt. Although motor adaptation helps, we find that as the strength of the feedback increases individual phenotypes cannot maintain the optimal operational regime in all environments, suggesting that diversity could be beneficial.

**Author Summary.** The biased random walk is a fundamental strategy used by many organisms to navigate their environment. Drift along the desired direction is achieved by reducing the probability to reorient whenever conditions improve. In the chemotaxis system of *Escherichia coli,* this is accomplished with a sensory module that implements negative integral feedback control, the output of which is relayed to the flagellar motors (the actuators) by a response regulator to control the probability to change direction. The proper dynamical coupling between sensor and actuator is critical for the performance of the random walker. Here, we identify an optimal regime for this coupling that maximizes drift velocity in the direction of the gradient in multiple environments. Our analysis reveals that feedback of the behavior onto the system in steep gradients can constrain individual cell performance, by causing bi-stable behavior that can trap cells in non-chemotactic states. These limitations are inherent in the biased random walk strategy with integral feedback control, but can be alleviated if the output of the pathway adapts, as recently characterized for the flagellar motors in *Escherichia coli*.

## Introduction

*Escherichia coli* cells navigate their environment by alternating straight runs with direction-changing tumbles to perform a random walk. During a run, the flagellar motors spin counterclockwise (CCW) and propel the cell at constant speed in one direction, which changes slowly due to rotational diffusion. Runs are terminated when one or more motors start rotating clockwise (CW), which causes the cell to tumble [1-3]. Cells are able to bias their random walk toward favorable conditions using a two-component signal transduction pathway that detects changes in signal intensity during runs and modulates the probability to tumble accordingly, resulting in extended runs in the desired direction and net drift velocity in the direction of the gradient [4].

The sensory module of the chemotaxis pathway (Figure 1A) consists of large clusters of receptor proteins that bind signal molecules to modulate rapidly (< 0.1s) the activity of an associated histidine kinase, CheA [5-7]. The high gain of the receptor cluster is coupled to negative integral feedback control [8-10], mediated by slow (~1-30 seconds) methylation and demethylation of the receptors by CheR and CheB, respectively [11-13]. This allows the receptors to adapt to a constant background signal while maintaining sensitivity over a wide range of concentrations [14,15]. For example, when cells are stimulated with a step of aspartate, the activity of the receptors returns nearly precisely to its pre-stimulus level after a transient response (Figure 1B first line). While precise adaptation does not hold when receptors become saturated, adaptation with a precision above 80% has been measured for many relevant signals within the micromolar range [16]. Precise adaptation is an important feature of bacterial chemotaxis because it provides robustness by implementing a "maximin" strategy that guarantees at least minimum chemotactic performance in any environmental condition [17].

The activity of the sensory module is relayed through a diffusible response regulator CheY to the flagellar motors, which act as the actuator, (Figure 1A). When phosphorylated by CheA, CheY-P binds to the motor subunit FliM and increases the probability of the motor to switch from CCW to CW [18]. Fast dephosphorylation of CheY-P by the phosphatase CheZ ensures rapid transfer of information from the sensor to the actuators.

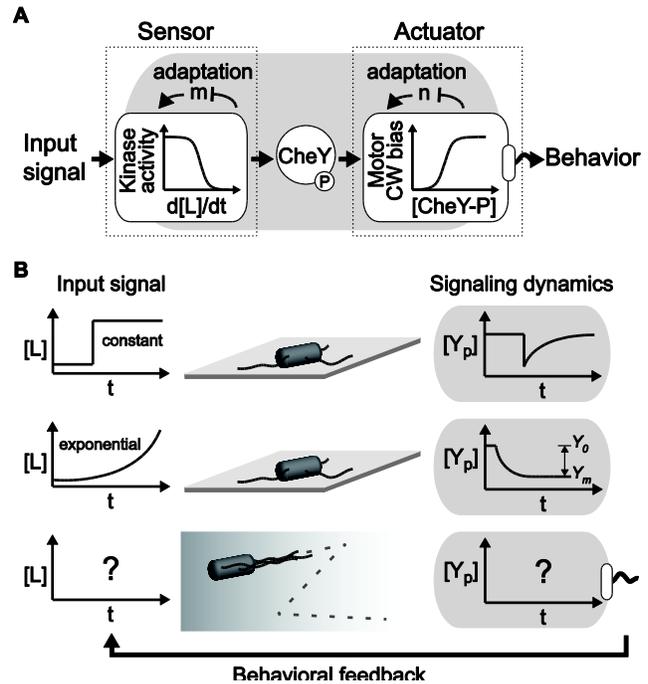

**Figure 1. Dynamical coupling between the sensor and the actuator in the bacterial chemotaxis system. A.** The bacterial chemotaxis system is composed of a sensor module (receptor-kinase complexes) and an actuator module (flagellar motors) coupled through the phosphorylated form of CheY. Both modules are ultra-sensitive and adapt to their respective input signals. Maintaining the output of the sensor within the right range relative to the actuator is critical for chemotaxis performance. **B.** Diagrams of the CheY-P concentration response to different signals. First line: when cells are immobilized onto a slide, a step stimulus of attractant (e.g. methylaspartate) causes a sudden decrease in CheY-P concentration followed by a slower adaptation. Because of the negative integral feedback architecture of the sensor module, CheY-P adapts back to its pre-stimulus level, the adapted CheY-P concentration, $Y_0$. Second line: when immobilized cells are exposed to an exponential ramp in time of the same stimulus, the system, which is log sensing, experiences a constant "force" and adapts towards an operational CheY-P concentration, $Y_m$, lower than the adapted level $Y_0$. This deviation of CheY-P activity from $Y_0$ to $Y_m$ changes the coupling between sensor and actuator. Third line: when cells are swimming in a gradient of attractant, their biased random walk causes them to climb the gradient. The average drift velocity of the cell up a chemical gradient affects the average input signal experienced by the cell. This creates a feedback of the behavior onto the input signal, which in turn can significantly affect the operating concentration of CheY-P and thus the coupling between sensor and actuator.

The CW bias of the flagellar motor, which defines the tumbling probability [2], is a sensitive function of the CheY-P concentration (Hill coefficient > 10, Figure 1A) [19,20]. The capability of the system to maintain the CheY-P concentration within the tight dynamic range of the motor CW bias response function (Figure 1A) is often used to investigate robustness to fluctuations in protein concentrations and receptor activity [21,22]. An important underlying assumption is that performance is maximized when the motor converts small variations in CheY-P into large changes in CW bias.

However, recent experiments and theory suggest that the coupling between sensor and actuator is more complex than previously thought. First, the flagellar motors partially adapt to persistent stimulus [23,24]. Second, the motor CW bias response to CheY-P is steeper than previously reported, further restricting the dynamic range of the motor response to CheY-P fluctuations [20]. Finally, in exponential ramps of chemoattractant, the CheY-P concentration reaches a dynamical equilibrium, $Y_m$, hereafter called *operational* CheY-P, distinct from the *adapted* CheY-P concentration, $Y_0$, that the cell maintains in constant uniform environments [25,26] (Figure 1B second line). For each of these three findings, the characterization of the internal dynamics of the signaling pathway was performed on immobilized cells using experimentally controlled input signals. However, for cells swimming freely in chemical gradients, the input signal dynamics are determined by the chemotactic response of the cell, creating a feedback of the behavior onto the input signal [27] (Figure 1B third line). Because of this behavioral feedback, it remains unclear how the multiple time scales of the system, from signal detection to motor response, ultimately determine chemotactic drift performance.

Here, we use analytical models and stochastic simulations of individual cells to examine the consequences of these new observations for our understanding of the bacterial chemotaxis strategy. Clonal populations of chemotactic *E. coli* grown in homogeneous conditions exhibit significant cell-to-cell phenotypic variability, with adaptation times ranging from 1 to 30 seconds [28-31], and motor clockwise bias ranging from 0.1 to 0.4 [3,31]. Therefore, we consider how different combinations of adaptation times and motor clockwise biases, which define a cell's phenotype, affect individual cell chemotactic drift velocity in different environments. In a phenotypically diverse population, different phenotypes of that population may perform best in different environments.

Focusing on how information transfer from sensor to actuator affects chemotactic performance, we analyze the dynamical relationship between the operational regime of CheY-P, $Y_m$, and the drift velocity, $V_D$, as a function of the phenotype and gradient steepness. We show that there is a unique operational regime of the sensor with respect to the motor that maximizes drift velocity in the direction of the gradient by maximizing the contrast between runs up and down the gradient, and not by maximizing the CW bias response. We characterize the performance trade-off faced by individual cells with different combinations of phenotypic parameters (such as, adapted CheY-P concentration, $Y_0$, receptor adaptation time, $\tau$, and cell resting tumble bias, $TB_0$).

# Results

**Maximizing contrast in run durations rather than CW bias response maximizes chemotactic drift velocity.**

Previous studies have examined how *E. coli* chemotactic drift velocity along a one dimensional gradient depends on the adaptation time [25,32], the shape of the response function of the sensory module [17,33,34] and also behavioral feedback [27]. Instead, we first focus on how the coupling between the sensors and actuators by the response regulator CheY-P (Figure 1A) affects the chemotactic performance of individual cell phenotypes. What CheY-P concentration maximizes drift velocity of cells navigating exponential gradients of methyl-aspartate? We examine this question using stochastic simulations of individual cells and an analytical model.

Simulations were conducted using a standard model of the chemotaxis pathway in individual cells [2,15] as described in Methods. Receptor-kinase complex activity is modeled as an all-or-none response using quasi-equilibrium dynamics for fast ligand binding, chemoreceptor conformational changes, and phosphorylation cascade [15]. The slower (de)methylation kinetics follow simple negative integral feedback dynamics with adaptation rate $\tau^{-1}$. The flagellar motor is modeled as an inhomogeneous Poisson process that switches cell behavior between runs and tumbles with rates defined as a function of CheY-P concentration that varies in time, $Y(t)$. The parameters of the motor model are calibrated to recent experimental measurements [19,20,23,24]. While motor adaptation [23] is not included at first, its effects are analyzed later in the paper. During runs, a cell swims with constant speed $v = 20$ μm$^{-1}$ in a direction subjected to rotational diffusion (rotation diffusion constant, $D_r = 0.062$ rad$^2$s$^{-1}$ [1]). For simplicity the effects of multiple flagellar motors [2,35] or directional persistence [1] are not included but discussed in the Discussion section. Hence, in this model, motor clockwise bias (CW) and cell tumble bias (TB) are the same. We consider cells containing only Tar receptors and use methyl-aspartate as the ligand. Our results readily extend to more complex receptor cluster configurations.

Three-dimensional trajectories of individual cells were simulated as described in [2] for various cell phenotypes, which are characterized by the receptor adaptation times $\tau$ and adapted CheY-P concentrations $Y_0$, in gradients of chemoattractant of different steepness, $g$. Following previous studies [25-27], we used exponential gradients ($L(x) = L_0 e^{gx}$) so that cells experience an approximately constant "force" from the attractant field, as the chemotaxis system is a fold-change detector (Eq. (2) below). This makes it possible to define a steady-state drift velocity, making the problem analytically tractable. The performance of each cell phenotype, which is defined by a unique adapted CheY-P concentration ($Y_0$) and receptor adaptation time ($\tau$), in each gradient steepness ($g$) is defined as the drift velocity $V_D(Y_0, \tau, g)$ along the gradient direction calculated by averaging the velocity of 10,000 phenotypically identical cells over 4 minutes (Methods). The first simulations were done in a relatively shallow gradient $g^{-1} = 5,000$ μm, with adaptation times of $\tau = 5$, 10, and 30 s, and adapted CheY-P concentrations spanning the range $Y_0 = 1$–4 μM.

Plotting drift velocity as a function of the adapted CheY-P concentration reveals that maximal drift velocity is not achieved for CheY-P concentrations in the linear range of the CW bias response curve, where fluctuations in CheY-P result in large changes in clockwise bias (Figure 2A). Instead, it occurs when adapted CheY-P is at the lower end of this curve (around 2.4 μM in Figure 2A).

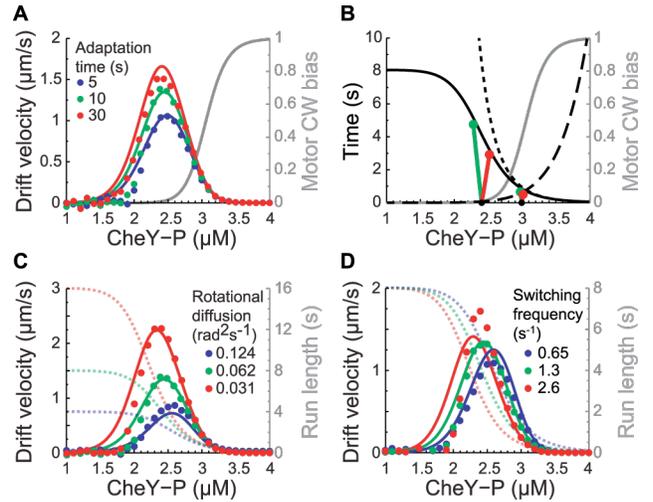

**Figure 2. Simulated and theoretical drift velocity $V_D$ in exponential gradient of aspartate $L_0 e^{gx}$. A.** $V_D$ as a function of the adapted CheY-P concentration $Y_0$, in a shallow gradient ($L_0 = 200$ μM and $g^{-1} = 5,000$ μm) for cells with adaptation times $\tau = 5$ (blue), 10 (green), and 30 seconds (red). $V_D$ is the average velocity of 10,000 identical cells between $t = 60$ and 300 seconds (dots: stochastic simulations; lines: analytical solution from Eq. (3); grey: motor CW bias response curve. **B.** Expected trajectories of CheY-P concentration $Y(F(t))$ for cells running in one dimension up (green) or down (red) in a gradient (integration of Eqs. (2) and (5), see SI text; $\tau = 30$ s, $g^{-1} = 5,000$ μm, $Y(F_i) = 2.4$ μM and 3 μM). Expected run, $\lambda_{R0}^{-1}$ (dotted line), and tumble, $\lambda_{T0}^{-1}$ (dashed line), durations as a function of $Y_0$. Expected run duration along a given direction $\tau_{R0} = (2D_r + \lambda_{R0})^{-1}$ (solid black line) is limited by rotational diffusion ($D_r = 0.062$ rad$^2$s$^{-1}$). Grey: motor CW bias. **C.** Same as A ($\tau = 10$ s) but with the rotational diffusion constant $D_r = 0.031$ (red), 0.062 (green), and 0.124 (blue) rad$^2$s$^{-1}$. Dotted lines: expected run duration in a given direction. **D.** Same as A ($\tau = 10$ s) but with the motor switching rate $\omega = 2.6$ (red), 1.3 (green), and 0.65 (blue) s$^{-1}$. Dotted lines: expected run duration in a given direction.

**Analytical model of the drift velocity as a function of CheY-P concentration.**

To understand the underlying reasons of this result, we derived an analytical relationship between CheY-P concentration and drift velocity along a one-dimensional gradient. For simplicity we used a one-dimensional analytical representation of bacterial chemotaxis in two or three dimensions [27,33,34,37,38]. In this framework, cells either go up or down the gradient or tumble and the effect of rotational diffusion can be represented as a jump process between runs up and runs down the gradient with transition rate $(d-1)D_r$, where $d$ represents the number of spatial dimension [37].

At quasi-steady state (for time scales longer than single run durations) and with no directional persistence (equal probability to run up or down the gradient), the drift velocity is proportional to the cell swimming speed $v$ times the difference between the expected run durations up $\langle t|+1\rangle_R$ and down $\langle t|-1\rangle_R$ the gradient divided by the total time including the time spent tumbling $\langle t\rangle_T$ [37]:

$$V_D = \frac{v}{d} \frac{\langle t|+1\rangle_R - \langle t|-1\rangle_R}{\langle t|+1\rangle_R + \langle t|-1\rangle_R + 2\langle t\rangle_T} \quad (1)$$

The only difference between $d = 2$ or $d = 3$ dimensions is a rescaling of the drift velocity and rotational diffusion (factor $d$ in the equations above; see SI text).

The expected run duration up or down the gradient is controlled by the cellular concentration of CheY-P, $Y$. This quantity is in turn a function of the free energy difference between the inactive and active receptor complexes, $F$, such that $F=\ln(\alpha/Y-1)$, where $\alpha$ is the gain of the phosphorylation cascade. The receptor activity follows simple spring-like dynamics around the adapted free energy difference $F_0$ with adaptation time $\tau$ (Methods, our Results still hold when considering asymmetric methylation/demethylation rates, see SI text and Figure S1):

$$\frac{dF}{dt} = -\frac{1}{\tau}(F-F_0) + sf \quad (2)$$

Here $s = \pm 1$ or 0 for cells running up, down the gradient, or tumbling, respectively. As the cell moves along a trajectory $x(t)$ it encounters different concentrations of the ligand $L(x(t))$. $f = vN\partial_x \ln[(1+L/K_i)/(1+L/K_a)]$ represents the magnitude of the change in free energy difference and depends on the local steepness of the gradient at the cell position. Here, $v$ is the speed of the cell when it is swimming, $N$ is the gain of the cooperative receptor system and $K_i$ and $K_a$ are the dissociation constants between ligand and receptors in the inactive and active conformation. In general, both $s$ and $f$ change as a function of time. For ligand concentrations $K_i \ll L \ll K_a$ we have $f \approx vN\partial_x \ln L$. If in addition, the gradient of ligand is exponential, $L=L_0 e^{gx}$, we see that $f \approx vNg$ becomes constant where $g$ represents the inverse length scale of the gradient. Therefore, in an exponential gradient the free energy difference of the receptors, $F$, tends to increase at the constant positive rate $+vNg$ when the cell swims up the gradient and to decrease at the constant negative rate $-vNg$ when the cell swims down the gradient. This in turn causes the CheY-P concentration to decrease (increase) when cells swim up (down) the gradient. The exact CheY-P concentration trajectories can be calculated by integrating Eq. (2) as a function of time for different initial condition $F_i$ (Figure 2B), while a cell is swimming up ($s=1$ green curve) or down ($s=-1$ red curve) a gradient.

The expected durations of a run, $\lambda_R^{-1}(Y)$, or a tumble, $\lambda_T^{-1}(Y)$, are plotted as a function of CheY-P concentration in Figure 2B (dashed lines; see definition in Methods). When a cell runs up or down the gradient, the rates of switching from one state to another change as a function of time and the direction of motion because they depend on the CheY-P trajectory. A run up or down the gradient can also be terminated by random reorientation from rotational diffusion with rate $(d-1)D_r$ [37]. Altogether, the rate at which a run is terminated by either rotational diffusion or a tumble is thus $\tau_R^{-1} = (d-1)D_r + \lambda_R(Y(F))$.

In a shallow gradient, $F$ deviates little from the adapted value $F_0$ and the adapted value of $\tau_R$ provides a good approximation of the expected run duration along a direction (black line in Figure 2B): $\tau_{R0} = \tau_R(F_0) = ((d-1)D + \lambda_{R0})^{-1}$, where $\lambda_{R0} = \lambda_R(F_0)$. When swimming up or down the gradient, CheY-P fluctuates (Figure 2B, green lines for up, red lines for down) and the run lengths are modulated approximately following $\tau_{R0}$ (black line and red and green circles in Figure 2B). According to equation (1), drift velocity is largest where the contrast between run duration up and down the gradient is the largest. Figure 2B reveals that this is the case where the slope of the expected run length as a function of CheY-P concentration is largest, which corresponds to the foot of the motor CW bias curve (Figure 2A) in agreement with the simulations. In contrast, for higher valued of CheY-P that are within the dynamic range of the CW bias response function, (e.g. $Y_0 = 3$ µM in Figure 2B) run durations up and down the gradient have a smaller contrast and longer tumble duration (dashed line Figure 2B), resulting in slower drift velocity.

In the limit of shallow gradients, Equation [1] can be linearized around the adapted values $F_0$ and $Y_0$ to obtain the drift velocity (Methods and SI Text):

$$V_D = \frac{\tau_{R0}'}{1 + \tau_{R0}/\tau} \frac{v(1 - TB_0)}{d} \int_0^\infty \frac{e^{-t/\tau_{R0}}}{\tau_{R0}} f dt$$
$$\cong \frac{\tau_{R0}'}{1 + \tau_{R0}/\tau} \frac{(1 - TB_0)v^2 Ng}{d} \quad (3)$$

Here, $TB_0 = \lambda_{R0}/(\lambda_{T0} + \lambda_{R0})$ represents the tumble bias of the cell as a function of the adapted CheY-P concentration $Y_0$ and the subscript $_0$ indicates that the rates $\lambda_R$ and $\lambda_T$ and $\tau_R' = d\tau_R/dF$ are all evaluated at the adapted state. The integral in Eq. (3) is the time-averaged input over the run durations, which in this approximation are exponentially distributed with characteristic time scale $\tau_{R0}$. For $K_i \ll L \ll K_a$ and exponential gradients, the rate of change of the free energy difference $s f \approx svNg$ is constant during a run. Equation (3) indicates that the drift velocity is proportional to the gradient steepness $g$ and the gain of the receptor cluster $N$. From Equation (3) we also obtain the chemotaxis coefficient of an individual cell phenotype: $\chi(Y_0, \tau) = V_D(Y_0, \tau)/g$.

Plotting the drift velocity as a function of $Y_0$ on top of the simulation results in Figure 2A shows that Equation (3) provides a good prediction of the drift velocity in shallow gradients and confirms that maximum velocity is reached for CheY-P values at the foot of the CW bias response curve.

In this linear regime, the optimal CheY-P concentration is only weakly dependent on the cell adaptation time and does not depend on the gradient steepness. The factor $\tau_{R0}'/(1 + \tau_{R0}/\tau)$ in Equation (3) encapsulates the relationship between drift velocity and the CheY-P concentration (or free energy difference $F$). For small adaptation times, $\tau \ll \tau_{R0}$, it increases linearly with adaptation time and is maximum where the slope $d \ln \tau_{R0}/dF_0$ is largest. For larger adaptation times, $\tau \gg \tau_{R0}$, this factor becomes $\approx \tau_{R0}' = Y_0(Y_0/\alpha - 1)d\tau_{R0}/dY$. Because the response function of the motor $\lambda_{R0} = \lambda_R(Y_0)$ (defined in Methods) is very steep (dashed line in Figure 2B), the slope $d\tau_{R0}/dY_0$ (slope of black line in Figure 2B) changes much faster as a function of $Y_0$ than $Y_0(Y_0/\alpha - 1)$.

Because rotational diffusion imposes an upper bound on the run length along a given direction [27,32], it determines, along with the motor parameters, the optimal range for CheY-P fluctuations. As rotational diffusion becomes smaller, cells are able to maintain their original direction for a longer time. The upper bound on the run length therefore becomes longer (dashed lines in Figure 2C) and the optimal CheY-P concentration becomes smaller (full lines in Figure 2C).

Changes in the switching frequency of the flagellar motor, ω, (see Methods) also affects the optimal CheY-P concentration. This becomes apparent when considering that the rate of switching from run to tumbles scales linearly with the switching rate of the flagellar motor, ω (see Methods). Therefore, the expected run duration of a cell scales like the inverse of ω. The result of this scaling is that for increasing values of ω the inflection point of the expected run length as a function of CheY-P shifts to lower values of CheY-P (dashed lines in Figure 2D). Thus, increasing the switching rate of the flagellar motors tends to decrease the optimal CheY-P concentration (full lines in Figure 2D). It also increases the maximum drift velocity that can be reached.

The analytical model of drift velocity in Eq. (3) is different from previous results [27] in two ways. First, it takes into account both the adaptation time and the tumbling state of the cell. Taking the limit $\tau \to \infty, TB_0 \approx 0$ in our model we recover the previous results. Second, in the previous study, the switching rate of the motor $\lambda_R$ was a steep function of kinase activity centered at the adapted kinase activity level, or equivalently, at the adapted CheY-P concentration $Y_0$. This means that changing $Y_0$ would also change the set point of the motor. However, the adapted CheY-P concentration and the set point of the motor are independent parameters. For this reason here we focused on the relationship between the set point of the motor and CheY-P activity. According to Eq. (3) the flagellar motor has its own sensitivity set point independent of the adapted CheY-P concentration of the sensory system $Y_0$ (see definition of $\lambda_R$ in Methods; motor adaptation [23,39] is considered below).

**A unique operational CheY-P concentration maximizes drift velocity for multiple gradients and adaptation times.**

Experiments have shown that when immobilized cells are exposed to an exponential ramp of methylaspartate, CheY-P activity reaches a new steady-state, $Y_m$, which is lower than its adapted activity, $Y_0$, because of the relatively slow adaptation rate of the system [9,26] (Figure 1B second line). When cells are swimming in an exponential gradient (Figure 1B third line), we expected a similar effect to take place because the average drift of an individual cell up the gradient will cause this cell to experience, on average, an exponential increase in ligand concentration as it makes its way up the gradient. While this effect should be minimal in a shallow gradient, it could become important in steep or rapidly changing gradients [27,40], especially for cells with longer adaptation times.

To investigate this issue we simulated cells swimming in a steeper exponential gradient ($g^{-1} = 1,000$ μm). After less than one minute of simulation, cell populations (10,000 replicate trajectories for each phenotype) reached a constant steady state drift velocity. We calculated the average ligand concentration that the cells encountered over time (Figure 3A). This reveals that the swimming cells experience an average exponential increase in ligand concentration over time. This average input is similar to the signal experienced by immobilized cells exposed to temporal exponential ramps (Figure 1B, second line) [9,26]. However, for the swimming cells the ramp rate is dynamically determined by the average drift velocity in the direction of the gradient (Figure 3A). Thus, for swimming cells the ramp rate depends on the feedback of the performance onto the input signal (Figure 1B, third line). Consistent with experimental results obtained with immobilized cells exposed to exponential ramps [26], the average CheY-P concentration in the swimming cells reaches a stable dynamical equilibrium, the operational value $Y_m$, after an initial drop from the adapted CheY-P concentration ($Y_0$) (Figure 3B).

The fact that the operational CheY-P concentration is not the same as the adapted CheY-P concentration implies that an optimal choice of adapted CheY-P must take into account this behavioral feedback. While a phenotype may for example have an adapted CheY-P concentration equal to the optimal concentration (~2.4 μM), during chemotaxis this level drops to an operational level lower than the optimum, hindering its performance (Figure 3B, solid black line). This effect is intensified when the adaptation time of the receptor cluster increases (Figure 3B, grey line). On the other hand, a phenotype with an adapted CheY-P concentration higher than the optimal concentration can approach the optimal operational CheY-P concentration as it reaches its steady-state drift velocity (Figure 3B, dotted line). The difference between $Y_0$ and $Y_m$ grows larger as drift velocity or the receptor adaptation time increase (Figure 3C).

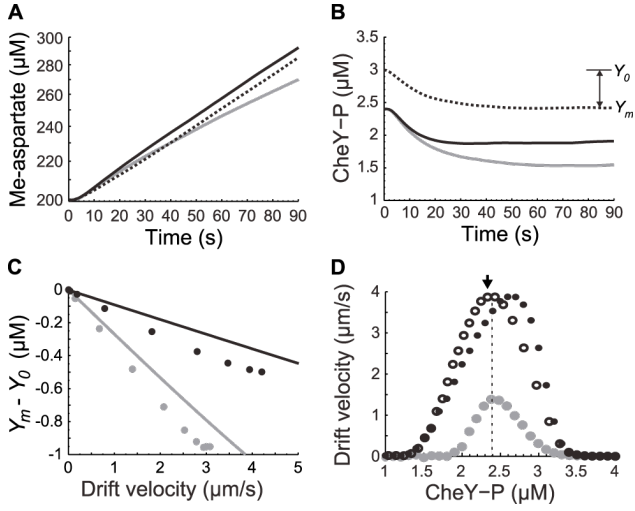

**Figure 3. Feedback of the behavior of cells swimming in exponential gradients onto the operational CheY-P concentration. A.** Temporal profiles of the average methyl-aspartate concentration encountered by cells swimming in a steep exponential gradient ($g^{-1} = 1,000$ μm). Different phenotypes are considered (solid black: $Y_0 = 2.4$ μM, $\tau = 10$ s, solid gray: $Y_0 = 2.4$ μM, $\tau = 30$ s, dotted black: $Y_0 = 3$ μM, $\tau = 10$ s) (the y-axis is on a log scale). **B.** Corresponding average CheY-P concentration as a function of time in these same cells **C.** Magnitude of the drop in average CheY-P activity (difference between adapted and operational CheY-P concentrations $Y_m - Y_0$) as a function of the drift velocity. Two different adaptation times are considered (black: $\tau = 10$ s, grey: $\tau = 30$ s). The gradient is the same gradient as in panel A. Dots are averages over 10,000 stochastic simulations for populations with different adapted CheY-P concentrations ($Y_0 > 2.4$ μM in both cases). Lines are from Eq. (4). **D.** Drift velocity $V_D$ as a function of adapted CheY-P concentration, $Y_0$ (filled circles), and operational CheY-P concentration, $Y_m$ (open circles) in stochastic simulations (average over 10,000 replicates for each circle, $\tau = 10$ s). $Y_m$ is instantaneous CheY-P concentration averaged over the population while drifting between $t = 60$ and $300$ s). Two exponential gradients of methyl-aspartate are considered ($g^{-1} = 1,000$ μm (black), 5,000 μm (grey)). Black arrow: cell population in blue in Figure 4C.

To determine whether the adapted or the operational CheY-P concentration is the primary variable that controls the average drift velocity in exponential gradients, we simulated cell populations with different adapted CheY-P concentrations and calculated their respective operational CheY-P concentrations. In a steep gradient ($g^{-1} = 1000$ μm), the optimal adapted CheY-P increased to ~2.7 μM compared to ~2.4 μM in a shallow gradient (Figure 3D). However, the optimal operational CheY-P concentrations for steep and shallow gradients are identical (Figure 3D). This suggests that a unique operational CheY-P concentration maximizes drift velocity in multiple gradients.

The situation is different when the feedback is strong. In this case the signaling pathway fluctuates around the operational values $F_m$ and $Y_m$, rather than the adapted values $F_0$ and $Y_0$. Therefore, we need to update the analytical model to describe the drift velocity, $V_D$, as a function of $F_m$. If we linearize the drift velocity equation around $F_m$ rather than $F_0$ we obtain an equation identical to Eq. (3) but with the subscript $_0$ replaced by $_m$ and $\tau_{Rm}$, $\tau'_{Rm}$, $\lambda_{Rm}$, $\lambda_{Tm}$, and $TB_m$ now functions of $F_m$. Knowing how $V_D$ depends on $F_m$ is not enough to calculate the drift velocity. We also need an equation that describes how $F_m$ depends on $V_D$. To model the effect of a constant drift velocity along the chemical gradient on the activity of the receptor cluster, we can expand Eq. (2) around $F_m$ and solve for quasi-steady state:

$$F_m = F_0 + V_D\ \tau\ N\ \frac{d}{dx}\ln\left(\frac{1+L(x(t))/K_i}{1+L(x(t))/K_a}\right) \quad (4)$$
$$\approx F_0 + V_D \tau N g$$

This expression quantifies the deviation between the operational free energy difference $F_m$ and the adapted free energy difference $F_0$ as a function of the drift velocity and is consistent with the results of our simulations (Figure 3C) and [27]. Equation (4) also makes clear that the strength of the feedback depends on adaptation time, the receptor cluster gain $N$, and the steepness of the gradient. Behavioral feedback strongly affects performance because it moves $Y_m$ away from the optimal operating point relative to the motor. This, in turn, affects the capability of the motors to best use the information carried by CheY-P.

By explicitly taking into account the effect of the behavioral feedback onto the coupling between the operating regime of CheY-P and the motor, Eqs. (3) (with $_{0\to m}$) and (4) extend previous studies [17,25,27,32-34,37] and reveal new possible dynamical regimes for the biased random strategy as shown below.

**Strong behavioral feedback can push the system through a bifurcation creating two possible chemotactic states for some cell phenotype: a fast drift state and a trapped state.**

For a given phenotype ($Y_0$, $\tau$) and gradient length-scale, the steady state drift velocity is determined by the intersection of two curves (Figure 4A). The first curve (solid line in Figure 4A) describes how the drift velocity depends on the operational CheY-P concentration, $Y_m$. It is defined by Equation (3) (with $_{0 \to m}$) and its profile can be interpreted as follows. For very low values of CheY-P the cell never tumbles. Thus, the cell diffuses equally in all directions and the net drift along the gradient is zero. For high values of CheY-P, the cell tumbles all the time so drift is zero as well. In between these two extremes, drift velocity is maximized for a specific value of the operational CheY-P concentration. However, $Y_m$ is not an independent variable. As we showed above (Eq. (4)), because of the feedback the behavior onto the input, the operational CheY-P concentration is itself a function of the drift velocity (which can also be written as: $Y_m = \alpha / (1 + e^{\tau \, N \, g \, V_D} (\alpha / Y_0 - 1))$ ). This equation defines the dashed line in Figure 4A, which intersects the horizontal axis at $Y_0$. Because each line in Figure 4A defines a relationship between $V_D$ and $Y_m$, the intersection between the two lines fully determines the drift velocity and the operational CheY-P concentration (black circle in Figure 4A) for a given phenotype and gradient.

When the feedback is weak ($\tau N g$ small, i.e. short adaptation time, small gain, or shallow gradient), the operational CheY-P concentration only exhibits a weak dependency on drift velocity and there is only one steady-state solution (intersection). Therefore, an appropriate adapted CheY-P concentration could be selected to ensure that operational CheY-P concentration is approximately optimal at all times (Figure 4A).

When the feedback is stronger, drift velocity always acts as a negative feedback onto the operational CheY-P concentration. In contrast, the effect of the operational CheY-P concentration onto drift velocity depends on whether the operational CheY-P concen-tration is below or above the CheY-P concentration that maximizes drift velocity (Figure 4A). Below this concentration, the system obeys negative feedback dynamics, whereas above it, the system obeys positive feedback dynamics. This positive feedback loops combined with the non-linear decrease of the drift velocity as a function of the operational CheY-P concentration, which arise from the extreme sensitivity of the flagellar motor, can lead to bistability [41]. Indeed, for a stronger feedback (steeper gradient or longer adaptation time) the slope of the feedback curve (dashed line in Figure 4AB), which is proportional to $1/\tau N g$, decreases. Thus, for phenotypes with high enough adapted CheY-P concentration ($Y_0$ is the intersection of the dashed line with the horizontal ax-is), the two curves can intersect more than once (Figure 4B). In this case, a single phenotype can now experience three different chemotactic states. Two of these states, $Y_{m1}$ and $Y_{m2}$ (filled circles in Figure 4B), are stable and are separated by one unstable state (open circle in Figure 4B). For one of the stable solution, the drift velocity is nearly zero and $Y_{m1}$ is high and very close to the adapted CheY-P concentration. For the other stable solution, the drift velocity is large and $Y_{m2}$ is much smaller than the adapted CheY-P concentration.

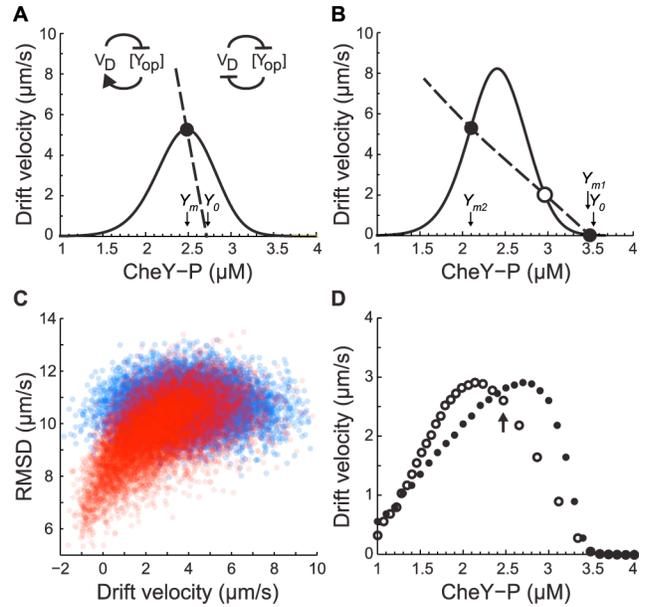

**Figure 4. Behavioral feedback can create a chemotactic "trap". A.** Analytical drift velocity, $V_D$ as a function of $Y_m$ (Eq. (3) with $_{0 \to m}$; solid line) and feedback of $V_D$ on $Y_m$ (Eq. (4), dashed line) ($\tau = 5$ s, $g^{-1} = 1,000$ µm, $Y_0 = 2.7$ µM). Steady state drift velocity ($Y_m = 2.48$ µM, black circle). **B.** Same as panel A, but for cells with longer adaptation time and higher adapted CheY-P ($\tau = 30$ s, $Y_0 = 3.5$ µM). $V_D$ has three possible steady states: two stable ($Y_{m2} = 2.1$ µM and $Y_{m1} = 3.49$ µM (black dots)), and one unstable ($Y_m = 2.97$ µM, white dot). **C.** Individual drift velocities (in the direction of the gradient) and root mean square displacements (perpendicular to the gradient) of two different populations of 10,000 simulated cells (blue: $\tau = 10$ s, $Y_0 = 2.6$ µM, red: $\tau = 30$ s, $Y_0 = 3$ µM). **D.** Average $V_D$ as a function of $Y_0$ (filled circles) and $Y_m$ (open circles) for cells with a long adaptation time ($\tau = 30$ s). Black arrow: cell population plotted in panel C (red).

This analysis suggests that an individual phenotype might experience two different chemotactic states with dramatically different performance: a fast drifting state and a "trapped" state. To find evidence of these two behaviors, we simulated two cell phenotypes (10,000 replicates for each phenotype) in a steep exponential gradient ($g^{-1}$ = 1,000 µm). One phenotype was predicted to operate closer to the bifurcation than the other (red and blue dots in Figure 4C, respectively). Although both phenotypes reached the same average operational CheY-P ($Y_m$ = 2.3 µM), cells with a phenotype closer to the predicted bifurcation point ($Y_0$ = 3µM, $\tau$ = 30s) exhibited a distribution of behavior (both drift velocity and diffusion) significantly skewed toward the "trapped" state (Figure 4C). Closer examination of the trajectories and CheY-P dynamics of individual cells in this simulation reveals that individual cells transition stochastically back and forth between the "trapped" and fast drifting state (Figure S2). For cases with higher feedback strength cells spend more and more time within the "trapped" state.

When the feedback is strong and the system becomes multistable, the average includes cells in both the "trapped" and high drift states. This phenomenon explains the decreased average drift velocities observed when adaptation time is increased (above 10 seconds) in a relatively steep gradient ($g^{-1}$ = 1,000 µm). It also explains the resulting shift of the best operational CheY-P concentration to lower concentrations (from ~2.4 to 2.1µM in Figure 4D), since for phenotypes with lower values of the adapted CheY-P only one stable state exists. Similar results are obtained when asymmetric methylation/demethylation rates of the receptors are taken into account (Figure S3).

**Motor adaptation partially alleviates the chemotactic "trap".**

Recent experiments have shown that the number of FliM monomers in the C-ring of the flagellar motor slowly (~minutes) adapts as a function of the CW bias, affecting both the steepness and the half-maximum CheY-P concentration of the CW bias motor response curve [24]. To examine the effect of motor adaptation on the relationship between CheY-P concentrations and drift velocity, we added motor adaptation to our stochastic model of an individual chemotactic cell by taking into account recent experimental data [20,23,24] (Methods). The resulting CW bias response curve of the adapted motor agrees well with both recent [20] and earlier [19] experimental measurements. In fact, it matches earlier experiments [19] better than a simple Hill function, suggesting that in these experiments the individual motors measured had adapted to the particular concentration of CheY-P expressed in the corresponding individual cells (Figure 5A; Methods).

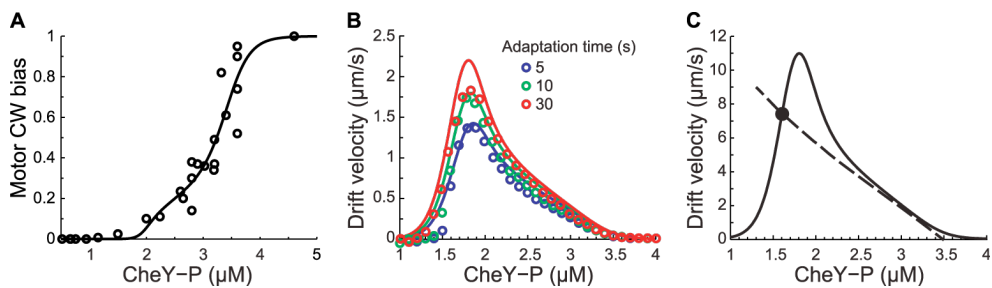

**Figure 5. Effect of motor adaptation on drift velocity $V_D$ in exponential gradients. A.** Motor CW bias response curve as function of CheY-P concentration when the motor is allowed to adapt (solid line) fitted to data from [19] (circles; derivation in Materials and Methods). **B.** Average drift velocity as a function of operational CheY-P concentration $Y_m$, in a shallow gradient. Same adaptation times and gradient steepness as Figure 2A. Lines: analytical solutions; circles: stochastic simulations (averages between $t$ = 10 and 15 min are used to calculate $V_D$ ($Y_m$)). **C.** Same as Figure 4B but with motor adaptation. The drift velocity has only one stable steady sate ($Y_m$ = 1.6 µM, black dot). Motor adaptation eliminated the other states present in Fig 4B.

Simulations of cells with motor adaptation in a shallow gradient ($g^{-1}$ = 5,000 µm) show that motor adaptation changes the shape of the drift velocity curve as a function of operational CheY-P, especially at high CheY-P concentrations (compare Figures 5B and 2A). These results are predicted by the analytical model (Eq. (3) with $_{0 \rightarrow m}$) once modified to include motor adaptation (Methods; lines in Figure 5B). Setting the adapted activity of the motor for a given CheY-P concentration to lower or higher CW biases gives qualitatively equivalent results (Figure S4 and S5).

How does the motor adaptation affect the bifurcation? In a steep gradient the behavioral feedback (Eq. (4)) must be taken into account (Figure 5C dashed line). Comparing Figure 5C and 4B we see that motor adaptation enable cells with high adapted CheY-P concentration to avoid the chemotactic trap improving performance (see Figure S6). This should provide a selective advantage because it helps buffer the functional consequences of inevitable cell-to-cell variability in the adapted CheY-P concentration, by increasing the range of CheY expression levels that allows effective chemotaxis.

Motor adaptation also affects the optimal operational CheY-P concentration (compare Figures 5B and 2A), shifting it to lower concentrations. When cells are drifting up a gradient, CheY-P drops to the operational CheY-P, causing the CW bias to drop. With motor adaptation, the lower operating CW bias causes the motor to shift its sensitivity to a lower CheY-P concentration. We see again that maximal performance is reached for $Y_m$ at the bottom of the CW bias response curve of the now adapted motor (Figure 5A). However, the motor can only compensate partially for the shift in operational CheY-P concentration.

## An individual phenotype faces a performance trade-off in different gradients.

As long as the system does not undergo bifurcation, maximum drift velocity is achieved by having a long adaptation time while maintaining the operational concentration of CheY-P in the optimal range. Therefore, the optimal adapted CheY-P concentration depends on the gradient length-scale and the adaptation time (Figure 6).

In shallow gradients, the strength of the feedback is small, as is the difference between operational and adapted CheY-P. Thus, it is possible to select an adapted CheY-P concentration that will perform relatively well for multiple adaptation times (Figure 6A blue line). In steeper gradients, the feedback is stronger (Eq. (4)) and the difference between $Y_m$ and $Y_0$ grows larger with adaptation time. Maintaining the optimal operational CheY-P concentration requires a higher adapted CheY-P concentration (Figure 5A green and red). The bifurcation of the system imposes an upper bound on the range of $Y_0$ beyond which a portion of the cells spend a significant amount of time trapped into a non-optimal state even with motor adaptation (Figure 6A dashed lines).

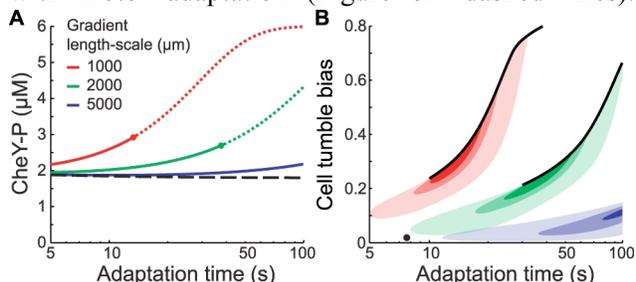

**Figure 6. Performance trade-off in bacterial chemotaxis. A.** Optimal adapted CheY-P concentrations $Y_0$ (solution of Eqs (4) with $_{0 \rightarrow m}$ and (5)) as a function of the chemoreceptor adaptation time in different exponential gradients ($g^{-1}$ = 1,000 (red), 2,000 (green), and 5,000 (blue) µm). Dots indicate when the maximal theoretical drift velocities cross the bifurcation point (dotted lines represent the inaccessible optimal state). The optimal operational CheY-P concentration $Y_m$ is identical for all gradient length scales (black dashed line). **B.** Contour plot of drift velocities as a function of adaptation time and the adapted cell tumble bias in different exponential gradients (same colors as A). 75%, 90%, and 95% contours of the maximal theoretical drift velocities for each gradient (colors intensities from light to dark). Black dot: the best cell phenotype that achieves equal relative drift velocities in all three gradients (60% of the maximal $V_D$ with $\tau$ = 7.5 $s$ and $TB_0$ = 0.044).

Therefore, the optimal adapted CheY-P concentration is a function of both receptor adaptation time and gradient length-scale, making it difficult for a single phenotype to maximize drift velocity in multiple environments (Figure 6A).

To characterize the resulting performance trade-off and map it to phenotypic space, we calculated the contours of drift velocity relative to its maximum in each environment, as a function of adaptation time $\tau$ and the adapted cell tumble bias (Figure 6B). In shallow gradients, cells benefit from a relatively long adaptation time and a low adapted CW bias. In steep gradients, cells benefit from a short adaptation time and a higher adapted CW bias. The best generalist phenotype can achieve at most 60% relative performance in all three gradients considered here. Motor adaptation, which was taken into account in generating Figure 6, alleviates only partially the tradeoff faced by single cells.

## Discussion

The adaptive response and feedback control of the receptor cluster play a critical role in the robustness of the chemotaxis system [8,10,15,17,33]. However, chemotactic performance also relies on the optimal operation of the flagellar motors, which directly control cell behavior. By focusing on how the CheY-P concentration affects the coupling between sensors and actuators, we revealed the existence of an *operational* regime for CheY-P concentration, which is distinct from the adapted CheY-P concentration, that maximizes drift velocity in a wide range of gradient length-scales and receptor adaptation times. Fluctuations around the best operational CheY-P concentrations maximize the contrast between run duration up and down the gradients. This occurs outside the most sensitive region of the CW bias response curve of the motor. Thus, chemotactic performance relies on maintaining the operational CheY-P concentration within bounds [21,22,42] around this optimal value.

The best operational CheY-P concentration is also determined by the cell rotational diffusion constant $D_r$, which imposes an upper bound on run durations in any particular direction (Figure 2C) [27,32]. In a more viscous environment or for longer cells, the lower rotational diffusion will result in a lower optimal operational CheY-P concentration. For an ellipsoid, rotational diffusion is inversely proportional the length of the major axis. Therefore, as cells grow the optimal range will shift to lower CheY-P concentrations. If the cell maintains a constant amount of CheY as it grew, the effective concentration of CheY-P would decrease, resulting in robust chemotactic performance during cell growth.

The switching frequency of the flagellar motors also affects the best operational CheY-P concentration. Higher switching frequencies tend to increase drift velocity while shifting the maximum to smaller CheY-P concentrations (Figure 2D). Therefore, the best operational CheY-P concentration is further away from the motor threshold. However, the range of CheY-P concentrations where the drift velocity is high becomes narrower (because the expected run length becomes a steeper function of CheY-P). This tends to increase the performance trade-off between different gradient length-scales. Thus, while selecting a higher switching frequency for the flagellar motors may improve performance of some phenotypes it may be detrimental for the population overall. Another important consideration is that the switching frequency is bounded by the speed at which the motor and associated flagella can switch confirmation [2,43].

Directional persistence (amount by which the swimming direction of a new run is biased towards the swimming direction of previous run) has been shown to affect chemotactic performance in climbing shallow gradients of attractants [1,36,44,45]. However, previous modeling and simulations efforts have been done using cells with non-optimal CheY-P concentrations (usually at 3 μM). In this regime, cells have a high tumbling rate, short run lengths, and low drift velocity. Directional persistence effectively reduces the reorientation rate of cells [45], which is equivalent to reducing the tumbling rate slightly. Therefore, directional persistence will shift the optimal CheY-P concentration to higher concentrations and improve the drift velocity of frequently tumbling cells [44]. On the other hand, when cells operate at or close to the optimal CheY-P concentration, the tumbling rate is low. Therefore, the run length in a given direction is terminated by rotational diffusion and not by tumbles. For optimal phenotypes, the relative effect of directional persistence on chemotactic drift is thus less important.

Previous studies have examined how the adaptation time affects chemotactic performance [12,25,27,32,46]. However, these studies only considered single values for the adapted CheY-P concentration (typically set to a CW bias of 0.5) and concluded that adaptation time should decrease as gradients get steeper to keep the operational CheY-P concentration within the dynamic range of the motor CW bias response. We found that, as long as the cell can maintain the optimal operational CheY-P concentration, longer adaptation time is better because it enhances input signal over the course of a run. However, long adaptation reinforces the feedback from the cell drift velocity on the system and can lead to undesirable bistability. Therefore, the bifurcation boundary imposes an upper limit on adaptation time as a function of the gradient length-scale. Interestingly, the distribution of tumble bias typically observed during exponential growth in single *E. coli* cells ranges from 0.1 to about 0.4 and not many cells are found that have higher tumble bias [31]. Selection for cells with tumble bias below 0.4 is consistent with our finding that the performance of cells with higher tumble bias will suffer from the existence of the "trapped" chemotaxis state.

Our results also provide a strong justification for the role of the recently-discovered flagellar motor adaptation. Indeed, we found that motor adaptation [23,39] plays a significant role in mitigating the behavioral feedback for cells with high tumble bias. When such feedback was included, cells with high tumble bias could escape the "trap" and gain access to a high drifting state in steep gradient. Our model also resolved an apparent contradiction between two sets of experimental measurements of the CW bias response of the flagellar motor as a function of CheY-P concentration. While one measurements reported a Hill coefficient of *n=10* [19], newer experiments reported a Hill coefficient of *n=20* [20]. In this paper we used the new value *n=20* and showed that the previous measurements are fitted with the same parameter value if one makes the reasonable assumption that the motors had had time to adapt before each individual cell measurement (Figure 5A).

Because the difference between the operational and adapted CheY-P concentrations depends on the strength of the behavioral feedback, which itself is proportional to gradient steepness, different adapted CheY-P concentrations and adaptation times are required to perform optimally in different gradients. Thus, in conditions where drift velocity is important, cells are faced with a performance trade-off. Even though motor adaptation was included, the best compromising phenotype over the gradient steepness considered in this study achieved at most 60% of the theoretical maximal drift velocity in all gradients. The observed cell-to-cell phenotypic diversity in adaptation time and adapted tumble bias [29,31] in an isogenic population may resolve the performance trade-off faced by single cells to improve the chance of survival of a unique genotype in complex or varying environments. In addition, the negative correlation between tumble bias and adaptation time observed by Park *et al.* in an isogenic population of *E. coli* [31], is consistent with our predictions about the most beneficial way to distribute phenotypes (Figure 6B).

At its core, the biased random walk relies on the dynamical control of the probability of reorientation. Overall, our analysis reveals limits to the use of negative integral feedback to control such strategy. Because the biased random walk strategy is used by many organisms, these results will inform our understanding of the constraints faced by other organisms as well.

## Materials and Methods

### Model and simulations

We used a standard model of bacterial chemotaxis [15] as described in [2]. For a cell following the trajectory $x(t)$, the output of the sensory module, the CheY-P concentration, is $Y(t) = \alpha/(1+e^F)$ where the free energy difference between inactive and ac-tive receptor complexes, $F = \varepsilon_0 + \varepsilon_1 m + N \ln((1+L/K_i)/(1+L/K_a))$, is a function of the methylation level, $m(t)$ and ligand concentration $L(\mathbf{x}(t))$. With $\alpha = 6$ μM, $\varepsilon_0 = 6$, $\varepsilon_1 = -1$, $N = 6$, and $K_i = 0.0182$ mM, $K_a = 3$ mM for methyl-aspartate and Tar receptors in the inactive and active conformation. When the cell is adapted to its environment, $F_0 = \varepsilon_0 + \varepsilon_1 m_0 = \ln(\alpha/Y_0 - 1)$. Adaptation mediated by methylation and demethylation of the receptors follows $dm/dt = -(m - \bar{m}(L))/\tau$, where $m - \bar{m}(L) = (F - F_0)/\varepsilon_1$. The methylation level $m$ is positive and bounded by the total number of methylation sites $m_{max} = 48$ available in a cooperative unit of receptors. The resulting adaptation dynamics fits recent experiments [26]. Cells switch between runs, $R$, and tumble, $T$, with rates $\lambda_{R,T} = \omega \exp[\mp G(Y(F))]$. The motor is modeled as a bistable system with switching frequency $\omega = 1.3$ s$^{-1}$ (unless otherwise stated) and free energy difference $G(Y) = \varepsilon_2/4 - (\varepsilon_3/2)(1+K/Y)^{-1}$ where $\varepsilon_2$ and $\varepsilon_3$ are non-dimensional constants that control the basal rate of switching of the motor when $Y=0$ and the degree of cooperativity of the motor, respectively. $K$ is the binding constant of CheY-P to FliM at the base of the motor. With $\varepsilon_2 = \varepsilon_3 = 80$, and $K = 3.06$ μM, this coarse-grained motor model fits well recent experimental measurements of CW bias (Hill coefficient 20) and switching frequency [20,23,24]. Motor adaptation is considered below.

**Linear expansion**

Eq. (2) follows by taking the time derivative of $F$ and using the relations from the previous section. Integration of Eq. (2) gives:

$$F(t, s, F_i) = F_0 + (F_i - F_0)e^{-t/\tau} + se^{-t/\tau}\int_0^t e^{u/\tau} f(u) du$$

The expected duration of a run along the direction $s = \pm 1$ is determined by the integral of the rate $\tau_R^{-1}$ of terminating a run along the direction $s$ by tumbling or because of rotational diffusion:

$$\langle t | \pm 1, F_i \rangle = \int_0^\infty e^{-\int_0^t \tau_R^{-1}(F(u, \pm 1, F_i)) du} dt \quad (5)$$

Because the average cell drift velocity in the direction of the gradient is determined by the contrast between expected run durations up and down the gradient (Equation (1)), the quantity of interest to calculate from Equation (5) is $\langle t|+1, F_i \rangle - \langle t|-1, F_i \rangle$. In a shallow gradient, the deviations from the adapted free energy difference $F_0$ are small. Considering only first order deviations $\Delta F = F - F_0$ around $F_0$ the change in free energy $|\Delta F / F_0|$ as a response to changes in ligand concentration is small and the inverse of the rate of run termination can be approximated by $\tau_R(F) \approx \tau_{R0} + \tau'_{R0}\Delta F$ where the mean run duration along a direction $\tau_{R0} = \tau_R(Y(F_0)) = ((d-1)D_r + \lambda_{R0})^{-1}$ and the gain $\tau'_{R0} = d\tau_R/dF$ are evaluated at $F_0$. Similar linear expansions are carried out for $\lambda_R$ and $\lambda_T$. Linear expansion of the free energy difference in Eq. (5) and integration by part gives:

$$\langle t|s, F_{iR}\rangle \cong$$
$$\tau_{R0}\left[1 - \lambda'_{R0}\int_0^\infty e^{-t/\tau_{R0}}\Delta F(t|s, F_i) dt\right] + O(\Delta F^2).$$

For tumble,

$$\langle t|F_{iT}\rangle \approx \int_0^\infty e^{-\lambda_{T0} t}\int_0^\infty \lambda_{R0} e^{-t_R/\tau_{R0}} dt_R dt + O(\Delta F)$$
$$= \tau_{R0} \lambda_{R0} / \lambda_{T0} + O(\Delta F).$$

Inserting in Eq. (1) and using the solution $F(t, s, F_i)$ we obtain the drift velocity to first order in $\Delta F$ (Eq. (3)).

**Motor adaptation**

The number of FliM molecules in the motor, $n$, is modeled as a binding and unbinding process with CW bias dependent rates [39]:
$$dn/dt = k_{on}(1-CW)/(1+\Delta n/(n_2 - n))$$
$$- k_{off} CW/(1+\Delta n/(n-n_1)).$$

The constants $k_{on}$ and $k_{off}$ define the rate of adaptation of the motor. $n_1$ and $n_2$ are the minimum and maximum FliM ring size that a motor can accommodate. $\Delta n$ is an effective half max parameter that guarantees that the effective rates of unbinding and binding to the motor go to zero when $n$ approaches $n_1$ or $n_2$. When $n$ changes it affects the steepness of the motor CW bias response, $CW = 1/(1+e^{2G})$, which in our case is controlled by $\varepsilon_3$ (see above). We used a simple linear relationship $\varepsilon_3 = \varepsilon_{3,1}(n-n_0) + \varepsilon_{3,0}$ where $\varepsilon_{3,1}$ is the slope and $n_0$ and $\varepsilon_{3,0}$ are the pre-stimuli level of the number of FliM and motor steepness, respectively. $k_{off}$= 0.025 s$^{-1}$, $n_1$ = 34, $n_1$ = 44 from [24]. We choose $\varepsilon_{3,0}$ = 80, $n_0$ = 36 to match the Hill coefficient of 20 measured for individual motor response curves [20], and fit $\Delta n$ = 2.74, $\varepsilon_{3,1}$ = 2.31 to reproduce [19] (Figure 5A). $k_{on}$ = 0.0063 s$^{-1}$ controls the CW bias that the motor adapts to (0.2 in this case, typical for wild type population of *E. coli* selected for swimming on agar plates [31]). At steady state, $dn/dt = 0$ defines $CW(n(\varepsilon_3))$ (Eq. [S20] in SI text). On the other hand, assuming quasi-equilibrium between the motor and operational CheY-P concentration $Y_m$, we have $CW(Y_m, \varepsilon_3) = 1/(1+e^{2G(Y_m,\varepsilon_3)})$. Solving the two equations gives $\varepsilon_3$ as a function of $Y_m$ from which we can calculate the drift velocity as Eq. (4) with motor adaptation (Figure 5).

## Acknowledgements

We thank N. Frankel, N. Olsman, A. Waite, J. Long, W. Pontius, and S. Zucker for helpful discussions. Simulations were performed at the Yale University High Performance Computing Center. This work is supported by the James McDonnell Foundation, The Paul G. Allen Family Foundation, and by NIGMS grant 1-R01-GM-106189-01.

# Supporting Text

**Mapping of the 2D and 3D problem onto 1 dimension: scaling of drift velocity and rotational diffusion**

Assume the cell swims along the direction $\vec{r}$, where $\vec{r}$ is a unit vector in $d = 2$ or 3 dimensions. Define $s(\theta) = \cos\theta$, where $\theta$ is the angle between the direction of motion and the direction of the gradient. Due to rotational diffusion [1], the directional of motion slowly drift away from the original direction during a run. The correlation function is $\langle \cos\theta \rangle(t) = \langle \vec{r}(t) \cdot \vec{r}_i \rangle = e^{-(d-1)D_r t}$, where $\vec{r}_i$ is the direction of motion at the beginning of the run. The survival probability distribution of a run along the direction $\vec{r}$ becomes,

$$F_R(t|\theta) = e^{-(d-1)D_r t} \cdot e^{-\int_0^t (\lambda_R(F(u,s_i,F_i))) du} \tag{S1}$$

where $F(t, s_i, F_i) = F_0 + (F_i - F_0)e^{-t/\tau} + e^{-t/\tau}\int_0^t e^{u/\tau} s_i f(u) du$ and $s_i = \cos\theta_i$ is the cosine of the angle between the direction of the signal gradient and the direction of the cell motion at the beginning of the run. In calculating the change in the free energy we neglected the higher order deviations in the angle of motion. The main effect of rotational diffusion is encapsulated in the first factor, $e^{-(d-1)D_r t}$.

At steady state, we linearize Eq. (S1) around the mean free energy $F_m$, which is on average lower than the resting free energy $F_0$.

$$F_R(t|s_i, F_i) = e^{-((d-1)D_r + \lambda_{Rm})t} \left[ 1 - \lambda'_{Rm} \int_0^t \Delta F(u, s_i, F_i) du \right] + O(\Delta F^2) \tag{S2}$$

$$F_T(t|s_i, F_i) = e^{-\lambda_{Tm} t} \frac{\lambda_{Rm}}{(d-1)D_r + \lambda_{Rm}} + O(\Delta F) \tag{S3}$$

where $\Delta F = F - F_m$. Integrating over time between 0 and infinity we get the expect run and tumble duration along the direction $s_i$

$$\langle t|s_i, F_i \rangle_R \cong \frac{1}{(d-1)D_r + \lambda_{Rm}} \left[ 1 - \lambda'_{Rm} \int_0^\infty e^{-((d-1)D_r + \lambda_{Rm})t} \Delta F(t, s_i, F_i) dt \right] + O(\Delta F^2) \tag{S4}$$

$$\langle t|s_i, F_i \rangle_T \cong \frac{\lambda_{Rm}/\lambda_{Tm}}{(d-1)D_r + \lambda_{Rm}} + O(\Delta F) \tag{S5}$$

Inserting Eqs. (S4) and (S5) in Eq. (1) the drift velocity is

$$V_D = v \frac{\int_{-1}^{1} \langle t|s_i, F \rangle_{iR} \, s_i P(s_i) ds_i}{\int_{-1}^{1} (\langle t|s_i, F_i \rangle_R + \langle t|s_i, F_i \rangle_T) P(s_i) ds_i} \tag{S6}$$

where $P(s_i) = 2\pi$ in 3D and $(1 - s_i^2)^{-1}$ in 2D. We are interested in the first order solution. Therefore only the zeroth order is needed for the denominator: $\frac{1 + \lambda_{Rm}/\lambda_{Tm}}{(d-1)D_r + \lambda_{Rm}} \int_{-1}^{1} P(s_i) ds_i$.

For the numerator we have to first order in $\Delta F$ and noticing that only the terms that are function of $s_i$ are not zero:

$$\frac{-\lambda'_{Rm}}{(d-1)D_r+\lambda_{Rm}}\int_{-1}^{1}ds_i\,s_i^2\,P(s_i)\int_0^\infty dt\,e^{-((d-1)D_r+\lambda_{Rm}+\tau^{-1})t}\int_0^t du\,e^{\frac{u}{\tau}}f(u)$$

$$=\frac{\tau'_{Rm}}{\tau_{Rm}}\int_{-1}^{1}ds_i\,s_i^2\,P(s_i)\left[-\frac{e^{-(\tau_{Rm}^{-1}+\tau^{-1})t}}{\tau_{Rm}^{-1}+\tau^{-1}}\int_0^t du\,e^{\frac{u}{\tau}}f(u)\Big|_0^\infty+\int_0^\infty dt\,\frac{e^{-\tau_{Rm}^{-1}t}}{\tau_{Rm}^{-1}+\tau^{-1}}f(t)\right]$$

The variation in $f(t)$ depends on the direction of motion

$$f(t)=vN\partial_x\ln[(1+L/K_i)/(1+L/K_a)]$$

The above relation shows that as long as $\int_0^t du\,e^{\frac{u}{\tau}}f(u)$ increases slower than $e^{(\tau_{Rm}^{-1}+\tau^{-1})t}$ (the ligand gradient is not steeper than $e^{e^{\tau_{Rm}^{-1}x/v}}$), the numerator becomes

$$\frac{\tau'_{Rm}}{\tau_{Rm}}\int_{-1}^{1}ds_i\,s_i^2\,P(s_i)\int_0^\infty dt\,\frac{e^{-\tau_{Rm}^{-1}t}}{\tau_{Rm}^{-1}+\tau^{-1}}f(t)$$

Thus, for $K_i\ll L\ll K_a$ and exponential gradients $f\approx vNg$ constant, the drift velocity in 2D/3D is

$$V_D\cong\frac{\tau'_{Rm}}{1+\tau_{Rm}/\tau}\frac{(1-CW_0)v}{d}\int_0^\infty\frac{e^{-t/\tau_{Rm}}}{\tau_{Rm}}f(t)\,dt\approx\frac{\tau'_{Rm}}{1+\tau_{Rm}/\tau}\frac{(1-CW_0)v^2Ng}{d} \tag{S7}$$

which is the same as Eq. (3) in the main text (after the subscript 0 has been replaced by $m$).

**Nonlinear solution**

All analytical curves in the paper (lines in Figures 2-5) use the linear approximation around $F_0$ and $F_m$ as described in the main text and Materials and Methods. Here we describe how to solve Eqs. (1)-(3) keeping the nonlinearity of the rates $\lambda_R(F)$ and $\lambda_T(F)$. Eq. (S16) can be integrated numerically to calculate the stopping time (red and green circles) of the red and green trajectories in Figure 2B. All other analytical curves in the paper (lines in Figures 2-6) use the linear approximation around $F_0$ and $F_m$ as described in Materials and Methods.

In the 1D representation, the equation for the free energy difference $F$ can be integrated to get $F(t,s,F_i)=(1-e^{-t/\tau})(F_0-sf\tau)+F_i e^{-t/\tau}$ where $s=\pm 1$ when the cell runs up or down the gradient and $s=0$ during tumbles. $F_i$ is the initial value. $f\approx vNg$ is the constant "force" exerted by the gradient.

Inverting we also get the time duration as a function of the free energy:

$$t(F,s,F_i)=\tau\log\left(\frac{F_i+sf\tau-F_0}{F+sf\tau-F_0}\right) \tag{S8}$$

At steady state the conditional probability densities of the duration $t$ of runs and tumbles are

$$P_{R\to T}(t|s,F_{iR})=\lambda_R(F(t,s,F_{iR}))e^{-\int_0^t((d-1)D_r+\lambda_R(F(u,s,F_{iR})))du} \tag{S9}$$

$$P_{R \to R}\left(t|s, F_{iR}\right) = (d-1)D_r e^{-\int_0^t \left((d-1)D_r + \lambda_R\left(F(u,s,F_{iR})\right)\right)du} \tag{S10}$$

$$P_{T \to R}\left(t|F_{iT}\right) = \lambda_T\left(F(t,0,F_{iT})\right) e^{-\int_0^t \lambda_T\left(F(u,0,F_{iT})\right)du} \tag{S11}$$

where $s = \pm 1$ and the first two probability densities correspond to runs that terminate into a tumble and into a run of opposite direction, respectively. $F_{iR}$ and $F_{iT}$ are the values of $F$ at the beginning of a run and a tumble, respectively. Noting that $dt = -\tau \dfrac{dF}{F + s\,f\,\tau - F_0}$ we also get

$$P_{R \to T}\left(F|s, F_{iR}\right) = P_{R \to T}\left(t(F,s,F_{iR})|s, F_{iR}\right)\frac{-\tau}{F + sf\tau - F_0} = \frac{-\tau \lambda_R(F)}{F + sf\tau - F_0} e^{\tau \int_{F_{iR}}^{F} \frac{(d-1)D_r + \lambda_R(F')}{F' + s\,f\,\tau - F_0}dF'} \tag{S12}$$

$$P_{R \to R}\left(F|s, F_{iR}\right) = P_{R \to R}\left(t(F,s,F_{iR})|s, F_{iR}\right)\frac{-\tau}{F + sf\tau - F_0} = \frac{-\tau (d-1)D_r}{F + sf\tau - F_0} e^{\tau \int_{F_{iR}}^{F} \frac{(d-1)D_r + \lambda_R(F')}{F' + s\,f\,\tau - F_0}dF'} \tag{S13}$$

$$P_{T \to R}\left(F|F_{iT}\right) = P_{T \to R}\left(t(F,0,F_{iT})|F_{iT}\right)\frac{-\tau}{F - F_0} = \frac{-\tau \lambda_T(F)}{F - F_0} e^{\tau \int_{F_{iT}}^{F} \frac{\lambda_T(F')}{F' - F_0}dF'} \tag{S14}$$

The probability density to have free energy $F_e$ at the end of a run and tumble cycle is then

$$P\left(F_e|F_{iR}\right) = \frac{1}{2}\sum_{s=\pm 1}\left(P_{R \to R}\left(F_e|s, F_{iR}\right) + \int_{-\infty}^{\infty} P_{T \to R}\left(F_e|F_{iT}\right) P_{R \to T}\left(F_{iT}|s, F_{iR}\right) dF_{iT}\right) \tag{S15}$$

At steady state we must have $P(F_e) = \int_{-\infty}^{\infty} P(F_e|F_{iR}) P(F_{iR}) dF_{iR}$ equal to $P(F_{iR})$, which given $P(F_e|F_{iR})$ defines $P(F_{iR})$. The average run and tumble durations are then

$$\langle t|s \rangle_R = \int_0^{\infty} dt\, t \int_{-\infty}^{\infty} dF_{iR} \left(P_{R \to R}(t|s, F_{iR}) + P_{R \to T}(t|s, F_{iR})\right) P(F_{iR}) \tag{S16}$$

$$\langle t|s \rangle_T = \int_0^{\infty} dt\, t \int_{-\infty}^{\infty} dF_{iR} P(F_{iR}) \int_{-\infty}^{\infty} dF_{iT} P_{T \to R}(t|F_{iT}) P_{R \to T}(F_{iT}|s, F_{iR}) \tag{S17}$$

We obtain the drift velocity

$$V_D = \frac{\frac{1}{2}\sum_{s=\pm 1} s \langle t|s \rangle_R}{\frac{1}{2}\sum_{s=\pm 1}\left(\langle t|s \rangle_R + \langle t|s \rangle_T\right)} \frac{v}{d} \tag{S18}$$

**Effect of asymmetric methylation/demethylation rates**

Experimental data shows the methylation/demethylation rates for receptor adaptation are asymmetric [2]. The rate of change of methylation catalyzed by CheR and CheB is usually described as

$$\frac{dm}{dt} = V_R \frac{1-a}{1-a+K_R} - V_B(a)\frac{a}{a+K_B} \qquad (S19)$$

where $V_R$ and $V_B(a)$ are the rates of methylation and demethylation; and $K_R$ and $K_B$ are the constants for each reactions. Experimentally, people found that the asymmetry of methylation/demethylation rates is not significant until $a > a_B$, where $a_B \approx 0.78$ measured in [2]. $V_B(a)$ is a piece-wise linear function:

$V_B(a) = V_{B,0}(1 + k_B \Theta(a-a_B)\frac{a-a_B}{1-a})$, where $\Theta(x)$ is a unit step function ($\Theta(x) = 1$ only if $x > 0$).

For most of the dynamic range of CheY-P level we are interested in $\Theta(a-a_B)$ will be zero and $V_B$ will be approximately constant. Thus variations in the rate of demethylation should not affect much drift velocity and optimal CheY-P level. To verify this, we implemented Eq. (S19) into our stochastic simulations of individual cells, with $K_R=0.43$, $K_B=0.3$ and $k_B=2.7$ [2]. To vary the adapted CheY-P level we varied $V_R$ and $V_{B,0}$ since their ratio determines the adapted CheY-P level. Given these definitions the effective adaptation time scale $\tau_{eff}$ obtained by linearizing equation (S19) reads $\tau_{eff} = \frac{a_0(a_0+K_B)(1-a_0+K_R)^2}{\left[a_0^2 K_R + ((1-a_0)^2 + K_R)K_B\right]V_R}$, where $a_0$ is the adapted activity of the receptor (corresponding to adapted CheY-P level $Y_0$ in this case). As shown in Fig. S1, the optimal CheY-P level in shallow gradient remains at the same position with respect to the motor response curve as in Fig. 2A. While the cells drifts in the steep gradient with slow methylation and demethylation rates, the behavior feedback will still push the system to a bifurcation (Fig. S2).

**Adapted motor response curve**

The motor adaptation is considered in this study by assuming that the number of FliM molecules in the motor changes as a function of the CW bias of the motor. At steady state, where $dn/dt = 0$, the relation between CW bias and number of FliM, $n$, is given by

$$CW(n) = \frac{(n-n_1+\Delta n)(n_2-n)\frac{k_{on}}{k_{off}}}{\left(1+\frac{k_{on}}{k_{off}}\right)(n_2-n)(n-n_1) + \Delta n\left[(n_2-n)\frac{k_{on}}{k_{off}} + (n-n_1)\right]} \qquad (S20)$$

Eq. (S20) together with the CW bias response function, $CW(Y_m, \varepsilon_3) = 1/(1+e^{2G(Y_m,\varepsilon_3)})$ and the linear relation between free energy $\varepsilon_3$ and the number of FliM $n$: $\varepsilon_3 = \varepsilon_{3,1}(n-n_0) + \varepsilon_{3,0}$, gives $\varepsilon_3$ as a function of $Y_m$. The adapted motor response curve $CW(Y_m, \varepsilon_3)$ is calculated then to fit the experimental data [3] with parameters $\Delta n$ and $\varepsilon_{3,1}$.

Note that for Figures 4 and 5 of the main text, $k_{on}$ was chosen as 0.0063 s$^{-1}$, so that the CW bias that the motor adapts to is 0.2, which is the average CW bias measured experimentally in wild type population of *E. coli* selected for swimming on agar plates [4]. We also examined what would happen if we changed the CW bias that the motor adapts to. For $k_{on} = k_{off} = 0.025$ s$^{-1}$ the effect of motor adaptation on the drift velocity curve is most visible for $Y_m$ between 2.5 and 3.5 µM (Fig. S3) whereas it is between 2 and 3 µM when $k_{on}$ = 0.0063 s$^{-1}$ (Fig. 4A). We also simulated the case where the CW bias that the motor adapts to is 0.05 (here $k_{on}$ = 0.0013 s$^{-1}$) which results in a flat region of the drift velocity curve as a function of $Y_m$ around the optimal operational CheY-P level (~2 µM) (Fig. S4). In this case, because $k_{on}$ is so small the adaptation time of the motor is very long and the motor does not reach steady state during the simulation. This explains the slight discrepancy with the analytical solution, which assumes steady state of the motor.

# Supporting Figures

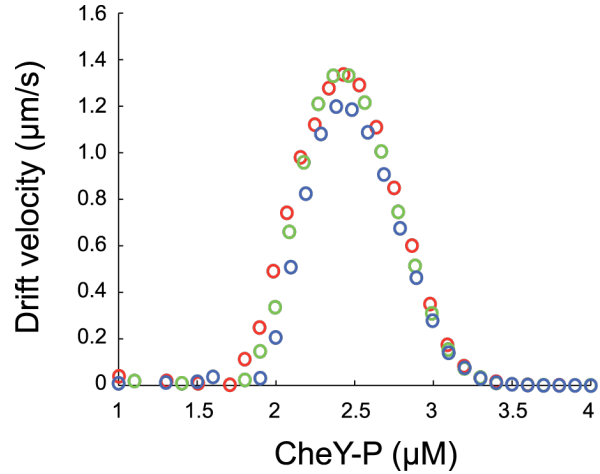

**Figure S1.** Effect of asymmetric methylation/demethylation rates on drift velocity $V_D$ in exponential gradient. Simulated drift velocity $V_D$ (average velocity of 10,000 cells between $t = 60$ and $300$ s) as a function of operational CheY-P concentration $Y_m$ in a shallow gradient ($L_0 = 200$ µM and $g^{-1} = 5,000$ µm) for cells with methylation rates $V_R = 0.1$ s$^{-1}$ (Red), $0.2$ s$^{-1}$ (Green), and $0.4$ s$^{-1}$ (Blue).

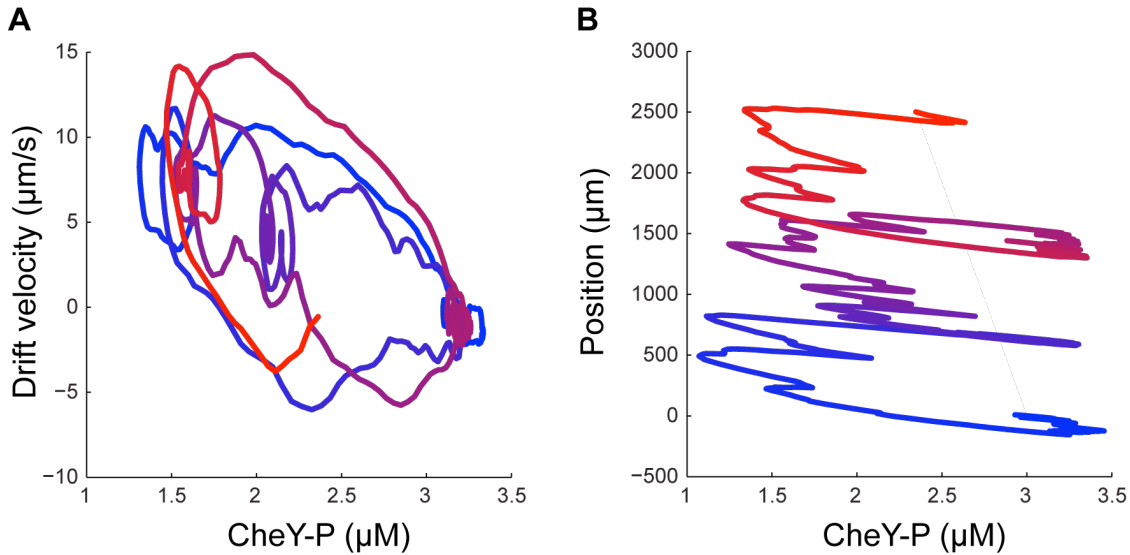

**Figure S2.** A simulated cell can transition in and out of the non-chemotactic state to reach the high drift velocity state when swimming in a steep gradient of methyl-aspartate ($g^{-1} = 1,000$ µm) illustrating the bi-stable behavior of this cell phenotype ($Y_0 = 3.0$ µM *and* $\tau = 30$ s). **A.** Single cell drift velocity as a function of its operational CheY-P concentration. When the cell escapes the "trapped" chemotactic state, characterized by a high CheY-P concentration, the behavioral feedback maintains an optimal CheY-P concentration and a high drift velocity. **B.** Cell position along the gradient as a function of its operational CheY-P concentration. The cell can escape the low drift velocity state and maintain a low CheY-P concentration when running up the gradient. On the other hand the cell can return to the "trapped" state after a long run down the gradient. The CheY-P concentration and drift velocity were calculated over a moving average window of 10 seconds. The time progression along the trajectory is indicated by the color of the stroke from blue to red.

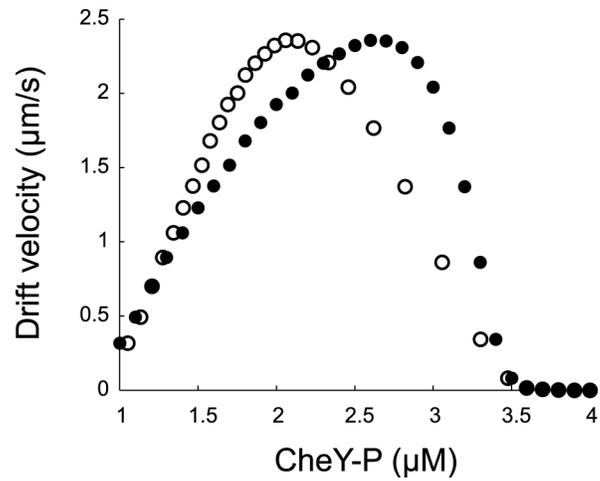

**Figure S3.** Effect of asymmetric methylation/demethylation rates on drift velocity $V_D$ in steep exponential gradient. $V_D$ from stochastic simulations (methylation rate $V_R = 0.1\,\text{s}^{-1}$) as a function of $Y_0$ (filled circles) and $Y_m$ (open circles) in exponential gradient of methyl-aspartate ($g^{-1} = 1{,}000\,\mu\text{m}$).

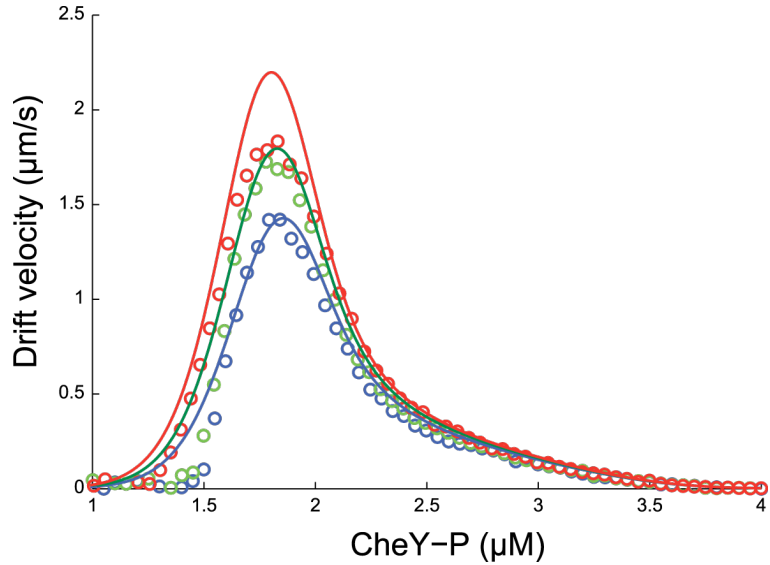

**Figure S4.** Drift velocity as a function of operational CheY-P when the rate of binding between FliM and the motor is $k_{on} = 0.025\,\text{s}^{-1}$. Circles are from simulations. Lines are from analytical solution. Everything is the same as in Figure 5A. The only difference is the value of $k_{on}$.

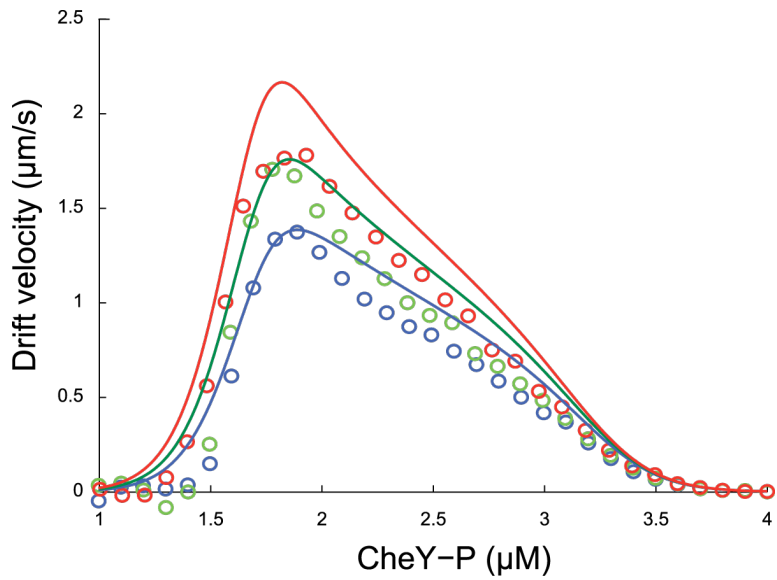

**Figure S5.** Drift velocity as a function of operational CheY-P when the rate of binding between FliM and the motor is $k_{on}$ =0.0013 s$^{-1}$. In this case, the motor does not adapt fast enough to reach quasi-steady state. The analytical solution (lines) makes the approximation that the system is at steady state.

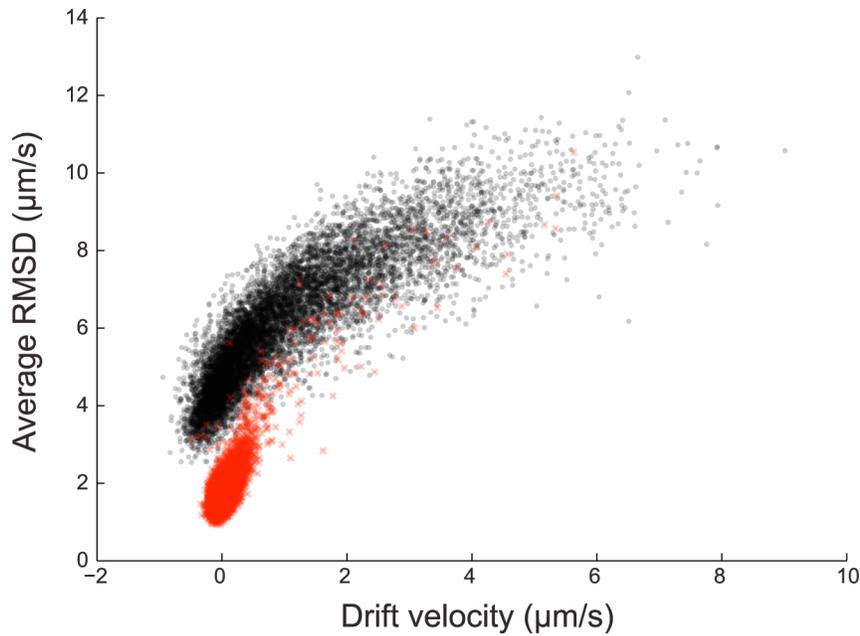

**Figure S6.** Scatter plot of individual drift velocities (in the direction of the gradient) and root mean square displacements (perpendicular to the gradient) of 10,000 simulated cells with motor adaptation (Black) and without motor adaptation (Red), with adaptation time $\tau$ = 30 s adapted CheY-P concentration $Y_0$ = 3.5 µM, gradient length scale $g^{-1}$ = 1,000 µm.